\documentclass[aps,prb,twocolumn,showpacs,preprintnumbers,amsmath,amssymb,floatfix]{revtex4-2}

\usepackage{graphicx}
\usepackage{longtable}
\usepackage{dcolumn}
\usepackage{bm}
\usepackage{color}
\usepackage[normalem]{ulem}
\usepackage[colorlinks=true, urlcolor=blue, linkcolor=blue, citecolor=blue]{hyperref}
\usepackage[all]{hypcap}
\newcommand{\beq}{\begin{equation}}
\newcommand{\eeq}{\end{equation}}
\newcommand{\bea}{\begin{eqnarray}}
\newcommand{\eea}{\end{eqnarray}}

\begin{document}


\title{Emergent room-temperature $d$-wave altermagnetism in antiferromagnetic superlattices}


\author{Ersoy \c{S}a\c{s}{\i}o\u{g}lu$^{1}$}\email{ersoy.sasioglu@physik.uni-halle.de}
\author{B\"{o}rge G\"{o}bel$^{1}$}
\author{Paul Richard Rei{\ss}mann$^{1}$}
\author{Ingrid Mertig$^{1}$}
\author{Samir Lounis$^{1}$}

\affiliation{$^{1}$Institute of Physics and Halle-Berlin-Regensburg Cluster of Excellence CCE, Martin Luther University Halle-Wittenberg, 06120 Halle (Saale), Germany}

\begin{abstract}
Despite recent experimental advances in
altermagnetism, realizing bulk $d$-wave
altermagnetism at room temperature remains
challenging.
Here, we demonstrate that $d$-wave altermagnetism
can emerge in superlattices based on conventional
spin-degenerate antiferromagnets.
Combining symmetry analysis, an effective
tight-binding model, and first-principles
calculations, we show that periodic stacking
can remove degeneracy-protecting symmetries
while preserving the combined fourfold
rotation and time-reversal symmetry
underlying $d$-wave spin splitting.
We identify 111-based, 122-based, and
Heusler-based altermagnetic superlattices
with nonrelativistic electronic spin
splittings reaching $261$~meV.
While the emergence of altermagnetism is
governed by crystallographic and magnetic
symmetries, the magnitude of the spin splitting
depends strongly on constituent selection
and superlattice periodicity.
Random-phase approximation calculations based
on first-principles exchange interactions yield
N\'eel temperatures above room temperature
for several superlattices, and the calculated
magnon dispersions exhibit characteristic
nonrelativistic $d$-wave splitting.
Our findings establish superlattice engineering
as a route to bulk room-temperature $d$-wave
altermagnetism with substantial electronic
spin splitting and robust magnetic order.
\end{abstract}

\maketitle

\begin{figure*}[!ht]
\centering
\includegraphics[width=0.8\textwidth]{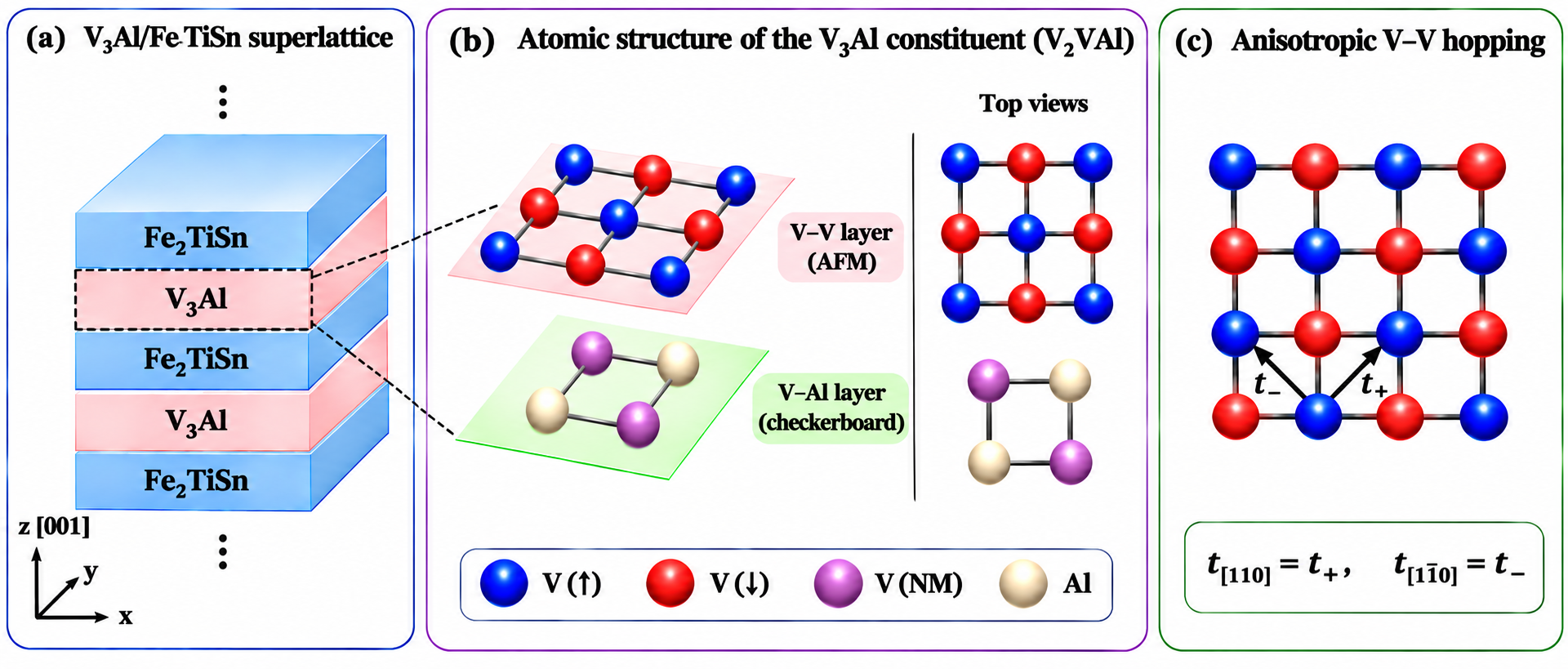}
\vspace{-0.1 cm}
\caption{\textbf{
Superlattice architecture and anisotropic hopping.}
(a) Schematic of the V$_3$Al/Fe$_2$TiSn
superlattice formed by periodic stacking
along [001].
(b) Atomic arrangement of the V$_3$Al
constituent, showing the antiferromagnetic
V--V layer and the adjacent V--Al
checkerboard layer in perspective and
top views. Blue and red spheres denote
oppositely magnetized V atoms, while
purple and cream spheres represent
nonmagnetic V and Al atoms, respectively.
(c) Schematic of anisotropic
intrasublattice V--V hopping, with
effective hopping amplitudes $t_+$ and
$t_-$ along the two in-plane diagonal
directions.
}
\label{fig0}
\end{figure*}

Altermagnetism has emerged as a distinct form of
compensated collinear magnetism characterized by
vanishing net magnetization and nonrelativistic,
momentum-dependent spin splitting of electronic
bands~\cite{Zunger,Smejkal2022PRX,Smejkal2022Landscape,
Song2025Review,Jungwirth2026Nature}.
Unlike conventional collinear antiferromagnets,
in which opposite-spin electronic states are
related by degeneracy-protecting symmetries,
altermagnets exhibit spin-split electronic bands.
Magnetic symmetry determines the momentum
dependence of the splitting, giving rise to
$d$-, $g$-, and $i$-wave spin-polarization
patterns~\cite{Smejkal2022PRX,
Smejkal2022Landscape,Jungwirth2026Nature}.
Recent spectroscopic experiments have established
bulk $g$-wave altermagnetism in
MnTe~\cite{Krempasky2024,MnTe2024ARPES,Lee2024MnTe}
and CrSb~\cite{Reimers2024CrSb,Ding2024CrSb,
CrSb2025ARPES,CrSb2026QuantumOscillations},
both of which exhibit magnetic ordering
temperatures above room temperature.
These advances have been complemented by
nanoscale imaging and control of altermagnetic
domains in MnTe~\cite{MnTe2024Imaging}.
In contrast, realizing bulk $d$-wave altermagnetism
with similarly robust magnetic order remains
challenging, despite its potential for
spin-dependent transport and spin-current
generation without net magnetization
~\cite{GonzalezHernandez2021,Jungwirth2026Spintronics}.

This challenge is illustrated by RuO$_2$ and
KV$_2$Se$_2$O, two extensively investigated
candidate $d$-wave altermagnets.
Rutile RuO$_2$ has played a central role in the
development of altermagnetism, with theoretical
predictions and experimental observations
supporting its altermagnetic interpretation
~\cite{GonzalezHernandez2021,Feng2022RuO2,
Fedchenko2024RuO2}.
However, its magnetic ground state remains
controversial, with recent experiments finding
no evidence of bulk altermagnetic spin splitting
or long-range antiferromagnetic order
~\cite{Liu2024RuO2,Hiraishi2024RuO2,
Keuss2024RuO2,Mizzi2026RuO2,Choi2026RuO2Review}.
First-principles calculations nevertheless predict
that altermagnetism can be stabilized in
ultrathin RuO$_2$ films through confinement
and structural relaxation~\cite{Brahimi2025RuO2}.
In KV$_2$Se$_2$O, spin-resolved photoemission
measurements revealed pronounced $d$-wave spin
splitting, whereas subsequent neutron diffraction
identified a G-type antiferromagnetic structure
incompatible with bulk altermagnetism
~\cite{KV2Se2O_ARPES,KV2Se2O_Neutron}.
Recent theoretical work suggests that these
observations may be reconciled by surface-induced
altermagnetism, in which surface termination
removes degeneracy-protecting symmetries of the
bulk magnetic structure
~\cite{Lange2026,JaeschkeUbiergo2026KV2Se2O}.
These findings motivate alternative
materials-design strategies for realizing
bulk $d$-wave altermagnetism.

One promising strategy is to start from
conventional antiferromagnets with high
magnetic ordering temperatures.
For example, V$_3$Al exhibits antiferromagnetic
order with a predicted N\'eel temperature
of approximately $600$~K
~\cite{Khmelevskyi2016V3Al}.
Layered 111- and 122-type compounds provide
further examples, including LiMnAs, NaMnAs,
and BaMn$_2$As$_2$
~\cite{Beleanu2013LiMnAs,Volny2022NaMnAs,
Singh2009BaMn2As2}, the latter exhibiting a
N\'eel temperature of approximately $625$~K.
Although these materials exhibit spin-degenerate
electronic bands in their conventional
antiferromagnetic states, their robust magnetic
order makes them attractive building blocks
for engineering room-temperature altermagnetism.

The potential of this strategy is illustrated
by recent studies of surface altermagnetism,
which demonstrate that breaking
degeneracy-protecting symmetries can induce
nonrelativistic spin splitting in otherwise
conventional antiferromagnets
~\cite{Sasioglu2026Surface,Lange2026,Hu2026}.
However, this effect is generally confined
to the vicinity of the surface.
This raises a fundamental question:
\emph{Can surface-induced altermagnetism
be transformed into a bulk phenomenon
through periodic stacking of suitable
materials?}
Here, we demonstrate that artificial
superlattices provide such a route by removing
degeneracy-protecting symmetries while preserving
the rotational symmetry underlying $d$-wave
spin splitting~\cite{Keshri2025,Bandyopadhyay2026}.
Using symmetry analysis, an effective
tight-binding model, and first-principles
calculations, we investigate [001]-oriented
111-based, 122-based, and Heusler-based
superlattices.
We establish the crystallographic and magnetic
symmetry conditions for bulk $d$-wave
altermagnetism and demonstrate how constituent
selection, magnetic stacking, and superlattice
periodicity control the resulting electronic
and magnetic properties.
We identify superlattices with nonrelativistic
electronic spin splittings reaching $261$~meV,
predicted magnetic ordering temperatures above
room temperature, and characteristic $d$-wave
magnon splitting.

\begin{figure*}[!ht]
\centering
\includegraphics[width=0.98\textwidth]{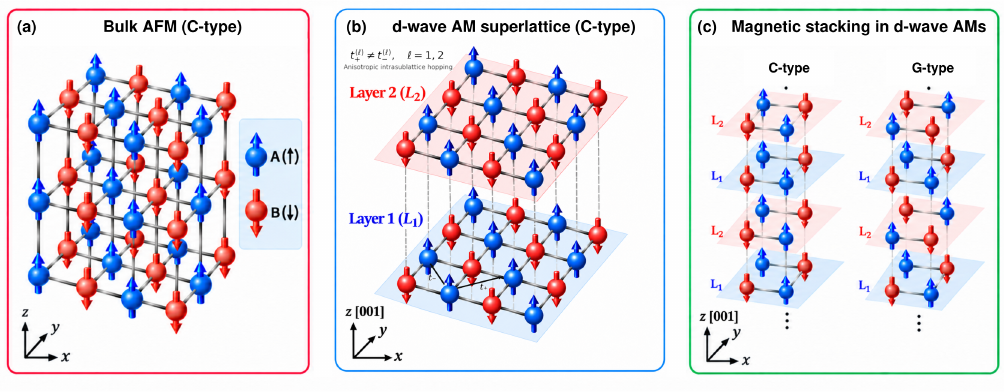}
\vspace{-0.1 cm}
\caption{\textbf{
Emergence of $d$-wave altermagnetism
through superlattice formation.}
(a) Schematic of a conventional bulk
C-type antiferromagnet with oppositely
magnetized sublattices $A(\uparrow)$
and $B(\downarrow)$.
(b) Schematic of a $d$-wave altermagnetic
superlattice formed by periodic stacking
of two different materials, $L_1$ and $L_2$,
along [001]. Superlattice formation removes
degeneracy-protecting symmetries while
retaining the fourfold rotation relating
the opposite-spin sublattices.
(c) C-type and G-type magnetic configurations
considered in the superlattices.
The in-plane antiferromagnetic arrangement
is repeated along [001] in the C-type
configuration, whereas the magnetic moments
reverse between successive layers in the
G-type configuration.
For clarity, only magnetic atoms are shown,
with blue and red spheres and arrows
denoting oppositely oriented magnetic moments.
}
\label{fig1}
\end{figure*}

\section*{Results}

We consider [001]-oriented superlattices
formed by alternating structurally compatible
antiferromagnetic and nonmagnetic constituents,
or two different antiferromagnetic materials.
As a representative example,
Fig.~\ref{fig0}(a) illustrates the
(V$_3$Al)$_{1/2}$/(Fe$_2$TiSn)$_{1/2}$
superlattice, whose electronic structure
will be discussed below.
Both constituents exhibit a layered atomic
structure along [001], with alternating
V--V and V--Al layers in V$_3$Al and
Fe--Fe and Ti--Sn layers in Fe$_2$TiSn.
For simplicity, Fig.~\ref{fig0}(b) shows
only the V$_3$Al constituent, in which
the magnetic V atoms form two sublattices
with antiparallel magnetic moments,
while the V and Al atoms in the adjacent
layer form a checkerboard arrangement.
In the superlattice, the
antiferromagnetic V--V layer is sandwiched
between two chemically inequivalent
checkerboard layers, namely V--Al and Ti--Sn.

The relative arrangement of the magnetic
V--V layer and the neighboring checkerboard
layers provides chemically distinct
environments for the two diagonal
intrasublattice hopping paths.
As illustrated for the V--Al layer in
Fig.~\ref{fig0}(b), these paths involve
hybridization with V and Al orbitals,
respectively, and are interchanged for
the opposite-spin sublattice.
The Ti--Sn layer on the opposite side
of the magnetic layer similarly contributes
to the direction-dependent orbital
hybridization.

In bulk antiferromagnetic V$_3$Al,
successive two-layer units of the type
shown in Fig.~\ref{fig0}(b) are rotated
by $90^\circ$ relative to one another
about the [001] axis, and the magnetic
V atoms exhibit conventional G-type
antiferromagnetic ordering.
In this bulk structure, the contributions
from the neighboring layers are related
by symmetry, and the combined
space-inversion and time-reversal
($\mathcal{PT}$) symmetry ensures
spin degeneracy throughout the
Brillouin zone.
In the superlattice, however, the chemical
inequivalence of the neighboring layers
removes this symmetry-imposed compensation,
allowing the effective hopping amplitudes
$t_+$ and $t_-$ to differ ($t_+\neq t_-$),
as illustrated in Fig.~\ref{fig0}(c).
The resulting sublattice-dependent hopping
anisotropy provides the microscopic origin
of the nonrelativistic $d$-wave spin splitting.

This mechanism is expected to be particularly
effective in ultrashort-period superlattices,
where every magnetic layer experiences
chemically inequivalent neighboring environments.
With increasing constituent thickness,
the local atomic and electronic environments
in the interior of the antiferromagnetic
constituent progressively approach those
of the corresponding bulk material,
leading to a suppression of the
spin-dependent hopping anisotropy.
The microscopic effects of the interface-induced
symmetry breaking are then expected to become
predominantly confined to the interfacial
regions.
Ultrashort-period superlattices thus offer
a route to extending an interface-induced
altermagnetic mechanism throughout a
three-dimensional periodic crystal.

\subsection*{Symmetry conditions for superlattice-induced
$d$-wave altermagnetism}
\label{sec:symmetry}

While the chemically inequivalent environments
of the magnetic layers provide a microscopic
mechanism for sublattice-dependent hopping
anisotropy, the emergence and momentum
dependence of altermagnetic spin splitting
are governed by the symmetry of the complete
magnetic superlattice.
The parent materials considered in this work
belong to three crystallographic families:
111-type compounds with space group
$P4/nmm$ (No.~129), 122-type compounds
with $I4/mmm$ (No.~139), and Heusler
alloys with $Fm\bar{3}m$ (No.~225).
Although these crystal structures are
centrosymmetric, the symmetries protecting
spin degeneracy depend on the magnetic order.
In antiferromagnets possessing combined
space-inversion and time-reversal symmetry,
$\mathcal{PT}$, opposite-spin states are
degenerate at the same crystal momentum,
\begin{equation}
E_{n\uparrow}(\mathbf{k})
=
E_{n\downarrow}(\mathbf{k}).
\label{eq:PT_degeneracy}
\end{equation}
Other magnetic symmetries, including
translation--time-reversal symmetry,
may also protect spin degeneracy in the
nonrelativistic limit, depending on the
magnetic structure.

Periodic stacking of chemically distinct
constituents along [001] modifies the
symmetry relations between magnetic
sublattices with antiparallel moments.
The altermagnetic superlattices identified
here possess the noncentrosymmetric
tetragonal space group $P4mm$ (No.~99),
which retains a fourfold rotational axis
along [001].
Superlattice formation removes certain
symmetries of the parent structures,
including inversion and the associated
$\mathcal{PT}$ symmetry where present.
However, inversion symmetry breaking
alone does not guarantee altermagnetism,
since spin degeneracy may still be
protected in the nonrelativistic limit
by a translation combined with time
reversal that relates magnetic sublattices
with antiparallel moments.

For the $d$-wave altermagnetic symmetry
considered here, the opposite-spin
sublattices are related by a fourfold
rotation combined with time reversal,
$C_4\mathcal{T}$.
In the nonrelativistic limit, using
the equivalence of $\mathbf{k}$ and
$-\mathbf{k}$ within each spin sector,
the corresponding spin-resolved band
energies satisfy
\begin{equation}
E_{n\uparrow}(\mathbf{k})
=
E_{n\downarrow}(C_4\mathbf{k}),
\label{eq:C4T_bands}
\end{equation}
where $C_4$ denotes a $90^\circ$
rotation about [001].
Defining the spin splitting as
\begin{equation}
\Delta E_n(\mathbf{k})
=
E_{n\uparrow}(\mathbf{k})
-
E_{n\downarrow}(\mathbf{k}),
\end{equation}
we obtain
\begin{equation}
\Delta E_n(C_4\mathbf{k})
=
-\Delta E_n(\mathbf{k}).
\label{eq:C4T}
\end{equation}
Thus, the spin splitting changes sign
under a $90^\circ$ rotation, reversing
the relative energy ordering of the
opposite-spin bands along the orthogonal
$X$--$\Gamma$ and $\Gamma$--$Y$
directions.
For the square-lattice geometry and
hopping anisotropy considered here,
the simplest lattice harmonic describing
this sign-changing spin splitting is
\begin{equation}
g_d(\mathbf{k})
=
\cos(k_xa)-\cos(k_ya),
\label{eq:symmetry_dwave_form}
\end{equation}
which has $d_{x^2-y^2}$ symmetry and
vanishes along $k_x=\pm k_y$.
Here, $k_x$ and $k_y$ refer to the
in-plane axes of the effective square-lattice
model, aligned with the two diagonal hopping
directions illustrated in Fig.~\ref{fig0}(c),
and $a$ denotes its lattice constant.

An additional consideration arises for
superlattices combining antiferromagnetic
and nonmagnetic constituents.
If the magnetic configurations in
successive superlattice periods are
related by a reversal of all magnetic
moments, the magnetic unit cell doubles
along [001].
The resulting translation--time-reversal
symmetry, together with the equivalence
of $\mathbf{k}$ and $-\mathbf{k}$
within each spin sector, restores
spin degeneracy throughout the magnetic
Brillouin zone in the nonrelativistic
limit.
Consequently, although $d$-wave spin
splitting may emerge in a magnetic
configuration constrained to the chemical
unit cell, it can disappear when the
magnetic periodicity of the ground
state is taken into account.
This behavior is demonstrated for the
122-based superlattices discussed in
the Supplementary Information,
highlighting the importance of the
magnetic ground state and its periodicity.

The selected $P4mm$ superlattices exhibit
the symmetry characteristics required
for $d$-wave altermagnetism in the
magnetic configurations considered.
Nevertheless, the crystallographic
space group alone does not determine
the presence of altermagnetic spin
splitting, and $P4mm$ is not a universal
requirement for $d$-wave altermagnetism.

While symmetry determines the allowed
momentum dependence of the spin splitting,
its magnitude is governed by the underlying
electronic structure and effective hopping
processes.
To elucidate the connection between the
sublattice-dependent hopping anisotropy
illustrated in Fig.~\ref{fig0}(c) and
the resulting $d$-wave spin splitting,
we introduce a minimal effective
tight-binding model in the following
subsection.
The model incorporates the effects of
the chemically inequivalent neighboring
layers through effective hopping
parameters.

\subsection*{Minimal effective tight-binding model for
superlattice-induced $d$-wave altermagnetism}
\label{sec:tb_model}

\begin{figure*}[!ht]
\centering
\includegraphics[width=1.02\textwidth]{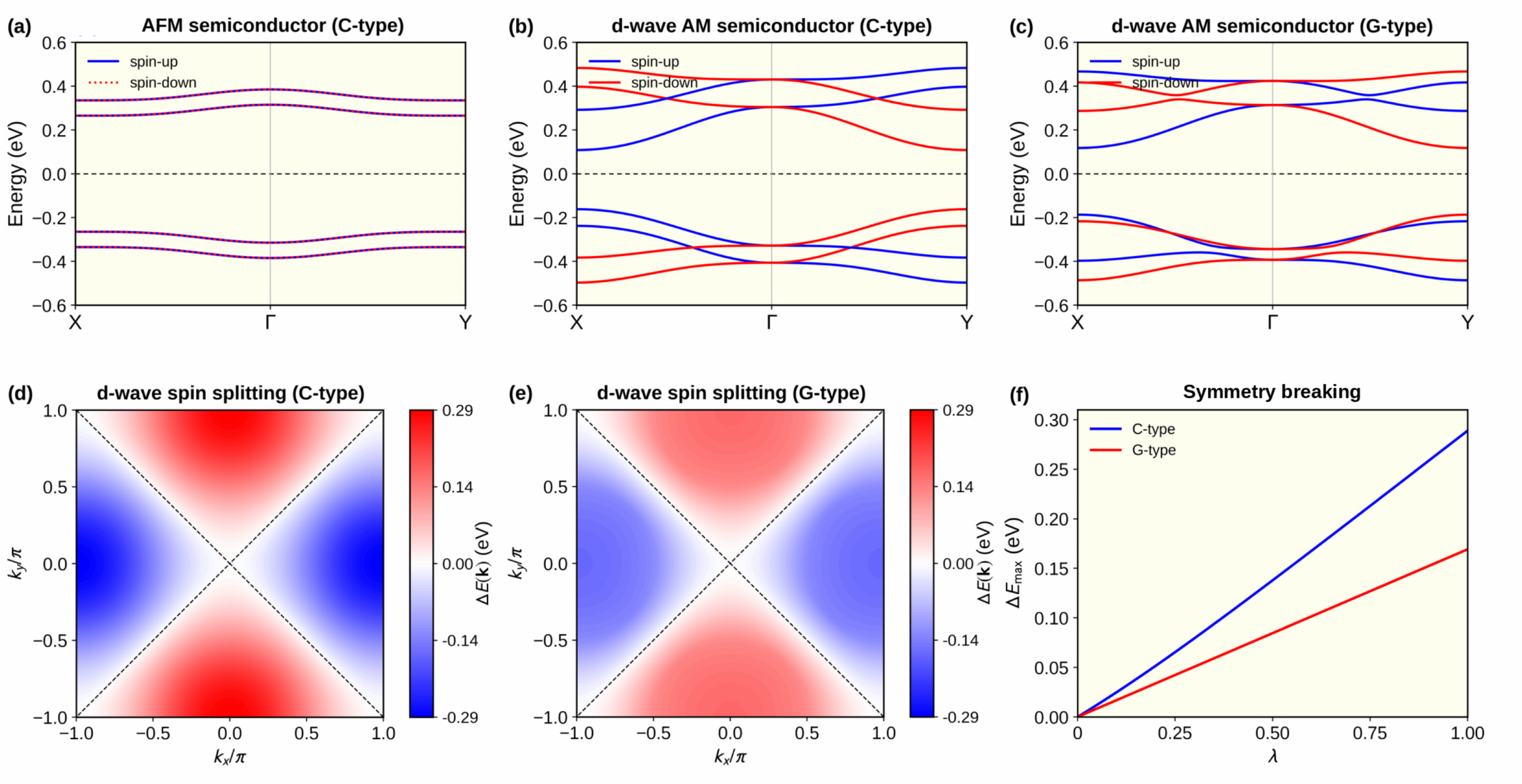}
\vspace{-0.8 cm}
\caption{\textbf{
Tight-binding description of superlattice-induced
$d$-wave altermagnetism.}
(a) Spin-degenerate band structure of the
C-type antiferromagnetic reference.
(b,c) Spin-resolved band structures of the
effective bilayer with C-type and G-type
magnetic stacking, respectively.
(d,e) Momentum-resolved spin splitting,
$\Delta E(\mathbf{k})=
E_{3\uparrow}(\mathbf{k})-E_{3\downarrow}(\mathbf{k})$,
for the third band (conduction band) in the
C-type and G-type configurations, respectively.
The sign reversal of the spin splitting under
a $90^\circ$ rotation and the diagonal nodal
lines are characteristic of $d$-wave altermagnetism.
(f) Evolution of the maximum spin splitting,
$\Delta E_{\max}=\max_{\mathbf{k},n}
|E_{n\uparrow}(\mathbf{k})- E_{n\downarrow}(\mathbf{k})|$,
as a function of the interpolation parameter
$\lambda$, connecting the spin-degenerate
antiferromagnetic limit ($\lambda=0$)
to the altermagnetic bilayer ($\lambda=1$).
For each magnetic configuration, the relative
magnetic orientation $(\eta_1,\eta_2)$ is
kept fixed throughout the interpolation.
The onsite energies, staggered exchange-field
magnitude $M_2$, and intrasublattice hopping
anisotropies are varied continuously, while
$M_1$, the hopping amplitude $t$, and the
interlayer hybridization $t_\perp$ remain fixed.
The model parameters and interpolation
scheme are provided in the Supplementary
Information.
}
\label{fig2}
\end{figure*}

We introduce a minimal effective tight-binding
model to illustrate the electronic mechanism
underlying superlattice-induced nonrelativistic
$d$-wave spin splitting. The model comprises
two inequivalent layers, $L_1$ and $L_2$,
stacked along [001], each represented by
a square lattice with two oppositely
magnetized sublattices, $A$ and $B$.
The layers represent effective electronic
environments within the superlattice rather
than isolated bulk constituents.
The anisotropic hopping processes are
motivated by the magnetic sublattice
geometry illustrated in Fig.~\ref{fig0}(c),
while the superlattice arrangement and
magnetic configurations are shown in
Fig.~\ref{fig1}.
Spin--orbit coupling is neglected, so
spin projection along the magnetic
quantization axis is conserved.

The Hamiltonian for layer $\ell$ is
\begin{equation}
H_{\ell}(\mathbf{k}) =
\varepsilon_{\ell}(\mathbf{k})\tau_{0}\sigma_{0}
+
\gamma_{\ell}(\mathbf{k})\tau_{x}\sigma_{0}
+
d_{\ell}(\mathbf{k})\tau_{z}\sigma_{0}
+
\eta_{\ell}M_{\ell}\tau_{z}\sigma_{z},
\label{eq:single_layer_tb}
\end{equation}
where $\tau_i$ and $\sigma_i$ are Pauli
matrices acting in the magnetic-sublattice
and spin spaces, respectively.
The term $\varepsilon_{\ell}(\mathbf{k})$
describes the sublattice-independent
dispersion, while $\gamma_{\ell}(\mathbf{k})$
represents hopping between opposite
magnetic sublattices.
The staggered exchange field has
magnitude $M_{\ell}$, and
$\eta_{\ell}=\pm1$ specifies its
relative magnetic orientation.

The symmetry-reduced electronic environment
is represented by anisotropic effective
same-sublattice hopping, schematically
illustrated in Fig.~\ref{fig0}(c).
We denote the two hopping amplitudes
by $t_+^{(\ell)}$ and $t_-^{(\ell)}$,
which are interchanged between the
opposite-spin sublattices under a
$90^\circ$ rotation.
For the two magnetic sublattices,
the hopping parameters satisfy
\begin{equation}
\begin{aligned}
t_{x,A}^{\mathrm{ss},(\ell)}
&=
t_{y,B}^{\mathrm{ss},(\ell)}
\equiv t_+^{(\ell)},\\
t_{y,A}^{\mathrm{ss},(\ell)}
&=
t_{x,B}^{\mathrm{ss},(\ell)}
\equiv t_-^{(\ell)}.
\end{aligned}
\label{eq:ss_hopping_relations}
\end{equation}
These relations describe anisotropic
hopping within each magnetic sublattice,
with the hopping amplitudes of the
opposite-spin sublattices related
by a $90^\circ$ rotation.

Defining the effective hopping anisotropy
as $\delta_{\ell}=t_-^{(\ell)}-t_+^{(\ell)}$,
the sublattice-dependent contribution
takes the form
\begin{equation}
d_{\ell}(\mathbf{k})
=
\delta_{\ell}
\left[
\cos(k_xa)-\cos(k_ya)
\right]
=
\delta_{\ell}g_d(\mathbf{k}),
\label{eq:d_wave_term}
\end{equation}
where $g_d(\mathbf{k})$ is the
$d$-wave form factor introduced
in Eq.~\eqref{eq:symmetry_dwave_form},
and $a$ is the in-plane lattice constant.

Without spin--orbit coupling,
the Hamiltonian separates into
independent spin sectors.
For $s=\pm1$,
Eq.~\eqref{eq:single_layer_tb}
becomes
\begin{equation}
H_{\ell s}(\mathbf{k})
=
\varepsilon_{\ell}(\mathbf{k})\tau_0
+
\gamma_{\ell}(\mathbf{k})\tau_x
+
\left[
d_{\ell}(\mathbf{k})
+
s\eta_{\ell}M_{\ell}
\right]\tau_z.
\label{eq:single_layer_spin_block}
\end{equation}
Assuming that
$\varepsilon_{\ell}(\mathbf{k})$
and $\gamma_{\ell}(\mathbf{k})$
are invariant under $C_4$, while
$d_{\ell}(C_4\mathbf{k})
=-d_{\ell}(\mathbf{k})$,
we obtain
\begin{equation}
\tau_x H_{\ell s}(\mathbf{k})\tau_x
=
H_{\ell,-s}(C_4\mathbf{k}).
\label{eq:tb_C4_relation}
\end{equation}
Here, $\tau_x$ exchanges the
magnetic sublattices.
This relation reproduces the
fourfold symmetry constraint
derived in the preceding subsection.

The corresponding eigenvalues are
\begin{equation}
E_{\ell s,\nu}(\mathbf{k})
=
\varepsilon_{\ell}(\mathbf{k})
+
\nu
\sqrt{
\gamma_{\ell}^{2}(\mathbf{k})
+
\left[
d_{\ell}(\mathbf{k})
+
s\eta_{\ell}M_{\ell}
\right]^2
},
\label{eq:single_layer_eigenvalues}
\end{equation}
where $\nu=\pm1$ labels the two bands.
For $\delta_{\ell}=0$, the spin
sectors are degenerate despite
the finite staggered exchange field.
For finite $\delta_{\ell}$,
the squared term contains the
spin-dependent contribution
\begin{equation}
2s\eta_{\ell}M_{\ell}\delta_{\ell}
\left[
\cos(k_xa)-\cos(k_ya)
\right],
\label{eq:cross_term}
\end{equation}
which generates nonrelativistic,
momentum-dependent spin splitting
when both the staggered exchange field
and hopping anisotropy are finite.
The splitting changes sign under
a $90^\circ$ rotation and vanishes
along $k_x=\pm k_y$, consistent
with the $d$-wave symmetry.
Although the splitting already
appears within an individual
effective magnetic layer, its
hopping anisotropy represents the
symmetry-reduced electronic environment
associated with superlattice formation.
Interlayer hybridization further
modifies the resulting band structure
and the magnitude of the splitting.

To describe the electronic coupling
between the two effective layers,
we introduce interlayer hybridization.
Using Pauli matrices $\rho_i$
in layer space, the bilayer
Hamiltonian reads
\begin{equation}
H(\mathbf{k})
=
P_1\otimes H_1(\mathbf{k})
+
P_2\otimes H_2(\mathbf{k})
+
t_{\perp}\rho_x\otimes\tau_0\otimes\sigma_0,
\label{eq:bilayer_hamiltonian}
\end{equation}
where $P_{1,2}=(\rho_0\pm\rho_z)/2$
project onto the two layers and
$t_{\perp}$ describes interlayer
hybridization.
Layer-dependent onsite energies,
exchange fields, and hopping
parameters account for the
inequivalent electronic environments
within the superlattice.

The interlayer coupling preserves
the fourfold spin-reversal symmetry
of the individual effective layers.
For two magnetic layers, parallel
($\eta_1=\eta_2$) and antiparallel
($\eta_1=-\eta_2$) orientations of
their staggered exchange fields
represent the C-type and G-type
configurations illustrated in
Fig.~\ref{fig1}(c), respectively.
Both configurations can exhibit
$d$-wave spin splitting, while
interlayer hybridization and
layer-dependent electronic parameters
influence its magnitude.

To illustrate the evolution from the
spin-degenerate antiferromagnetic reference
to the effective altermagnetic superlattice,
we introduce an interpolation parameter
$\lambda\in[0,1]$.
The layer-dependent onsite energies,
staggered exchange fields, and hopping
anisotropies are varied continuously
between their reference and superlattice
values, while the interlayer hybridization
$t_{\perp}$ is kept fixed.
Thus, $\lambda=0$ corresponds to the
spin-degenerate reference, whereas
$\lambda=1$ represents the fully developed
effective superlattice model.
The interpolation is phenomenological
and does not represent a microscopic
simulation of superlattice formation.

Figure~\ref{fig2} illustrates
the resulting electronic structure.
The symmetric C-type
antiferromagnetic reference
exhibits spin-degenerate bands
[Fig.~\ref{fig2}(a)].
Introducing anisotropic hopping
and layer inequivalence produces
$d$-wave spin splitting in both
C-type and G-type configurations
[Figs.~\ref{fig2}(b,c)].
For the chosen parameters,
the maximum splitting reaches
approximately $0.29$ and
$0.17$~eV, respectively.
These values illustrate the
dependence on electronic parameters
and magnetic stacking rather
than a universal hierarchy
between the configurations.
The momentum-resolved splitting
[Figs.~\ref{fig2}(d,e)]
changes sign under a $90^\circ$
rotation and vanishes along
the diagonal directions. 
Figure~\ref{fig2}(f) shows the
evolution of the maximum spin
splitting, $\Delta E_{\max}$,
with the interpolation parameter
$\lambda$.
The continuous increase in splitting
illustrates the role of the
symmetry-reduced electronic environment
in establishing the altermagnetic
band structure.

The model can also describe
the effective electronic coupling
between magnetic and nonmagnetic
layers by setting $M_2=0$.
This limit is relevant to
superlattices combining
antiferromagnetic and nonmagnetic
constituents, such as
LiMnAs/NaAlSi,
BaMn$_2$Ge$_2$/SrPd$_2$As$_2$,
and V$_3$Al/Fe$_2$TiSn,
while superlattices containing
two magnetic constituents,
such as LiMnAs/NaMnAs,
are represented by finite
exchange fields in both layers.

Although the effective model captures
the essential mechanism of
superlattice-induced $d$-wave
altermagnetism, the magnitude of the
spin splitting and the stability of
the altermagnetic state depend on
the material-specific electronic
structure and magnetic interactions.
To establish the realization of this
mechanism in realistic materials,
we now turn to first-principles
calculations of representative
superlattices.

\subsection*{Material realizations of superlattice-induced
$d$-wave altermagnetism}
\label{sec:materials}

Guided by the symmetry analysis and
the effective tight-binding model,
we investigate representative
111-based, 122-based, and Heusler-based
superlattices using density functional
theory (DFT).
These structural families provide
a broad range of antiferromagnetic
and nonmagnetic constituents with
compatible crystal structures and
diverse electronic properties,
allowing us to explore the proposed
mechanism across different chemical
compositions.
The superlattices are constructed
by periodic stacking of the
constituents along [001].
Throughout, $(A)_n/(B)_n$ denotes
a superlattice in which each constituent
has a thickness corresponding to
$n$ conventional bulk unit cells
along [001]; $n=1/2$ therefore
denotes half the corresponding
bulk unit-cell thickness.
The atomic stacking sequences
are provided in the Supplementary
Information.

\begin{table*}[t]
\centering
\caption{\textbf{Structural symmetry, electronic character, and magnetic properties of representative $d$-wave altermagnetic superlattices.}
The crystallographic space groups (SGs) of the bulk materials and corresponding [001]-oriented superlattices are listed together with their electronic character, nonrelativistic spin splitting, magnetocrystalline anisotropy energy (MAE), and calculated N\'eel temperature ($T_{\mathrm N}$). For metallic superlattices, $\Delta E_{\max}$ denotes the maximum spin splitting throughout the Brillouin zone among bands for which at least one spin branch crosses the Fermi level; the maximum need not occur at the Fermi energy. For semiconducting LiMnAs/NaMnAs, the splittings of the valence (VB) and conduction (CB) bands near the gap are reported separately. N\'eel temperatures are estimated within the classical random-phase approximation (RPA). All listed superlattices have $P4mm$ (No.~99) crystallographic symmetry. A dash indicates a value not reported for the corresponding configuration.}
\label{tab:superlattices}

\begin{tabular}{lcccccc}
\hline\hline
Superlattice
& Bulk SG
& SL SG
& Electronic character
& $\Delta E_{\max}$ (meV)
& MAE (meV/cell)
& $T_{\mathrm N}$ (K) \\
\hline
$(\mathrm{LiMnAs})_1/(\mathrm{NaMnAs})_1$ (G-type)
& $P4/nmm$ (129) & $P4mm$ (99)
& Semiconductor & 231 (VB), 38 (CB) & 1.28 & 285 \\
$(\mathrm{LiMnAs})_1/(\mathrm{NaMnAs})_1$ (C-type)
& $P4/nmm$ (129) & $P4mm$ (99)
& Semiconductor & 237 (VB), 63 (CB) & -- & -- \\
$(\mathrm{LiMnAs})_1/(\mathrm{NaAlSi})_1$
& $P4/nmm$ (129) & $P4mm$ (99)
& Metal & 261 & 0.64 & 352 \\
$(\mathrm{BaMn_2Ge_2})_{1/2}/(\mathrm{SrPd_2As_2})_{1/2}$
& $I4/mmm$ (139) & $P4mm$ (99)
& Metal & 87 & 0.35 & 1423 \\
$(\mathrm{V_3Al})_{1/2}/(\mathrm{Fe_2TiSn})_{1/2}$
& $Fm\bar{3}m$ (225) & $P4mm$ (99)
& Metal & 206 & 0.11 & 325 \\
$(\mathrm{V_3Al})_1/(\mathrm{Fe_2TiSn})_1$
& $Fm\bar{3}m$ (225) & $P4mm$ (99)
& Metal & 10 & 0.13 & 505 \\
\hline\hline
\end{tabular}
\end{table*}

Constituent selection was guided primarily by structural compatibility and magnetic character. The 111-type compounds LiMnAs, NaMnAs, and NaAlSi crystallize in $P4/nmm$ (No.~129), while the 122-type compounds BaMn$_2$Ge$_2$ and SrPd$_2$As$_2$ adopt $I4/mmm$ (No.~139). Within the Heusler family, we consider antiferromagnetic V$_3$Al and V$_2$TiSi, belonging to the 18-valence-electron class associated with the Slater--Pauling relation $M=Z-18$, together with nonmagnetic Fe$_2$TiSn, Fe$_2$VAl, and Ru$_2$VAl from the 24-valence-electron class associated with $M=Z-24$. Here, $M$ denotes the total magnetic moment per formula unit in units of $\mu_{\mathrm B}$, and $Z$ is the number of valence electrons. These materials allow us to investigate superlattices containing either two antiferromagnetic constituents or one antiferromagnetic and one nonmagnetic constituent.

For each constituent pair, the common in-plane lattice parameter was selected from the calculated equilibrium lattice parameters of the bulk compounds, followed by relaxation of the out-of-plane lattice parameter and internal atomic coordinates. Optimized structures, lattice parameters, local magnetic moments, and computational details are provided in the Supplementary Information. Table~\ref{tab:superlattices} summarizes the electronic and magnetic properties of the representative altermagnetic superlattices, all of which adopt the noncentrosymmetric tetragonal space group $P4mm$ (No.~99).

\begin{figure*}[!ht]
\centering
\includegraphics[width=1.0\textwidth]{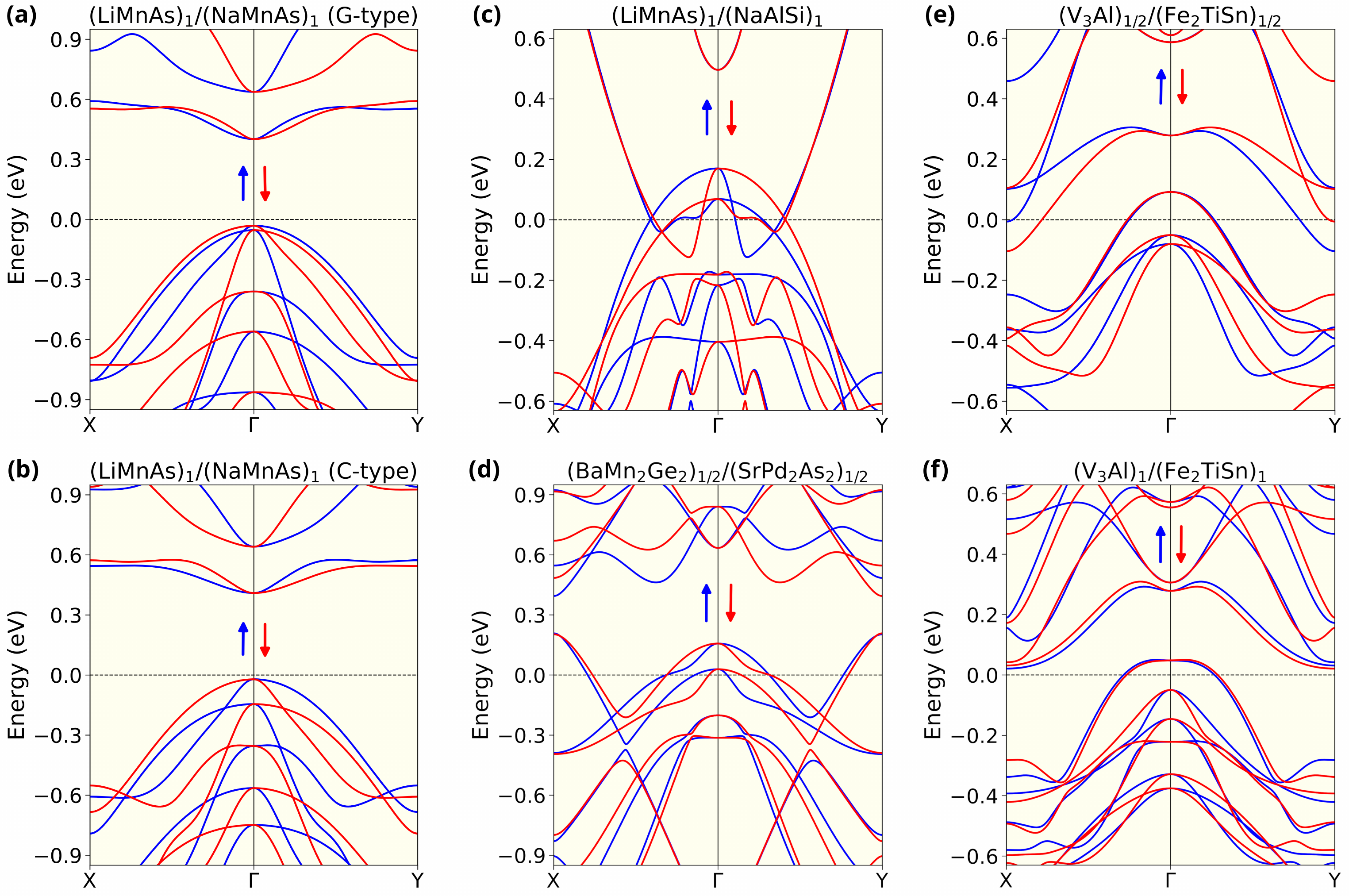}
\vspace{-0.4 cm}
\caption{\textbf{First-principles realization of $d$-wave altermagnetism in representative superlattices.}
Spin-resolved electronic band structures calculated without spin--orbit coupling along the $X$--$\Gamma$--$Y$ path.
(a,b) The 111-based $(\mathrm{LiMnAs})_1/(\mathrm{NaMnAs})_1$ superlattice in the (a) G-type magnetic ground state and (b) C-type configuration. Both are semiconducting, with direct band gaps of approximately $0.43$~eV.
(c) Metallic 111-based $(\mathrm{LiMnAs})_1/(\mathrm{NaAlSi})_1$.
(d) Metallic 122-based $(\mathrm{BaMn_2Ge_2})_{1/2}/(\mathrm{SrPd_2As_2})_{1/2}$.
(e,f) Metallic Heusler-based V$_3$Al/Fe$_2$TiSn superlattices with (e) $(\mathrm{V_3Al})_{1/2}/(\mathrm{Fe_2TiSn})_{1/2}$ and (f) $(\mathrm{V_3Al})_1/(\mathrm{Fe_2TiSn})_1$. Blue and red bands denote opposite spin channels.}
\label{fig3}
\end{figure*}

Figure~\ref{fig3} shows the spin-resolved electronic band structures calculated without spin--orbit coupling. The $X$--$\Gamma$--$Y$ path highlights the characteristic reversal of spin splitting between orthogonal momentum directions. The maximum splitting values are defined in Table~\ref{tab:superlattices}.

We first consider the 111-based $(\mathrm{LiMnAs})_1/(\mathrm{NaMnAs})_1$ superlattice, in which both constituents are antiferromagnetic. Total-energy calculations identify the G-type configuration as the magnetic ground state. The superlattice is semiconducting, with a direct band gap of approximately $0.43$~eV [Fig.~\ref{fig3}(a)]. Pronounced nonrelativistic $d$-wave spin splitting occurs near the band edges, reaching approximately $231$~meV in the valence bands and $38$~meV in the conduction bands.

We also examine the C-type configuration to assess the influence of magnetic stacking [Fig.~\ref{fig3}(b)]. The direct band gap remains approximately $0.43$~eV, while the valence-band splitting changes slightly to $237$~meV and the conduction-band splitting increases to $63$~meV. Thus, magnetic stacking affects the valence- and conduction-band splittings differently without changing the semiconducting character or $d$-wave symmetry.

Metallic $d$-wave altermagnetism also emerges in superlattices containing one antiferromagnetic and one nonmagnetic constituent. The 111-based $(\mathrm{LiMnAs})_1/(\mathrm{NaAlSi})_1$ superlattice [Fig.~\ref{fig3}(c)] exhibits a maximum nonrelativistic spin splitting of $261$~meV. The 122-based $(\mathrm{BaMn_2Ge_2})_{1/2}/(\mathrm{SrPd_2As_2})_{1/2}$ superlattice [Fig.~\ref{fig3}(d)] is likewise metallic, with $d$-wave spin splitting reaching approximately $87$~meV. Together with the LiMnAs/NaMnAs results, these calculations demonstrate that superlattice-induced $d$-wave altermagnetism can occur in both semiconducting and metallic systems.

To examine the influence of constituent selection and magnetic periodicity, we investigate additional 122-based superlattices (Supplementary Table S2 and Fig.~S2). The characteristic spin splitting calculated within the chemical supercell varies from approximately $3$~meV in $(\mathrm{BaMn_2As_2})_{1/2}/(\mathrm{BaNi_2As_2})_{1/2}$ to $177$~meV in $(\mathrm{YMn_2Ge_2})_{1/2}/(\mathrm{CaRh_2As_2})_{1/2}$, demonstrating a strong dependence on constituent selection.

However, spin splitting in the chemical supercell
does not necessarily imply altermagnetism in the
magnetic ground state. Unlike the LiMnAs/NaMnAs
superlattice, where both constituent layers are
magnetic, the AFM/NM superlattices require a
doubled magnetic unit cell for G-type coupling.
The $(\mathrm{YMn_2Ge_2})_{1/2}/
(\mathrm{CaRh_2As_2})_{1/2}$ and
$(\mathrm{SrCr_2As_2})_{1/2}/
(\mathrm{BaCo_2As_2})_{1/2}$ superlattices
favor C-type configurations, for which the
magnetic and chemical periodicities coincide
and the calculated splittings of 177 and
90 meV, respectively, are retained. 
In contrast, the other 122-based superlattices
favor G-type configurations. The resulting
translation combined with time reversal,
together with the relevant nonrelativistic
symmetries, restores spin degeneracy. This is illustrated by $(\mathrm{YCr_2Si_2})_{1/2}/(\mathrm{BaCo_2As_2})_{1/2}$, whose chemical-supercell splitting reaches $152$~meV but is absent in the complete magnetic structure. These results demonstrate that constituent selection influences the magnitude of the symmetry-allowed splitting, while the magnetic ground state determines whether global altermagnetism is realized.

Heusler-based superlattices provide another realization of the proposed mechanism. In $(\mathrm{V_3Al})_{1/2}/(\mathrm{Fe_2TiSn})_{1/2}$, periodic stacking reduces the crystallographic symmetry of the centrosymmetric cubic constituents to $P4mm$ (No.~99). The calculated band structure [Fig.~\ref{fig3}(e)] exhibits pronounced nonrelativistic $d$-wave spin splitting, reaching $206$~meV.

Not all Heusler-based superlattices exhibit altermagnetism. The shortest-period V$_3$Al/Ru$_2$VAl and V$_3$Al/Fe$_2$VAl superlattices retain centrosymmetric $P4/nmm$ (No.~129), with $\mathcal{PT}$ symmetry protecting opposite-spin degeneracy. Superlattices containing the two antiferromagnetic constituents V$_3$Al and V$_2$TiSi also remain spin degenerate. Their shortest-period and full-unit-cell structures adopt $P4/mmm$ (No.~123) and $P4/nmm$ (No.~129), respectively, with $\mathcal{PT}$ symmetry preserved in the magnetic state. The corresponding structures and band dispersions are presented in the Supplementary Information. These reference superlattices demonstrate that structural compatibility, electron count, and magnetic character alone are insufficient to establish altermagnetism; the atomic arrangement and complete magnetic symmetry must also be considered.

\begin{figure}[t]
\centering
\includegraphics[width=0.49\textwidth]{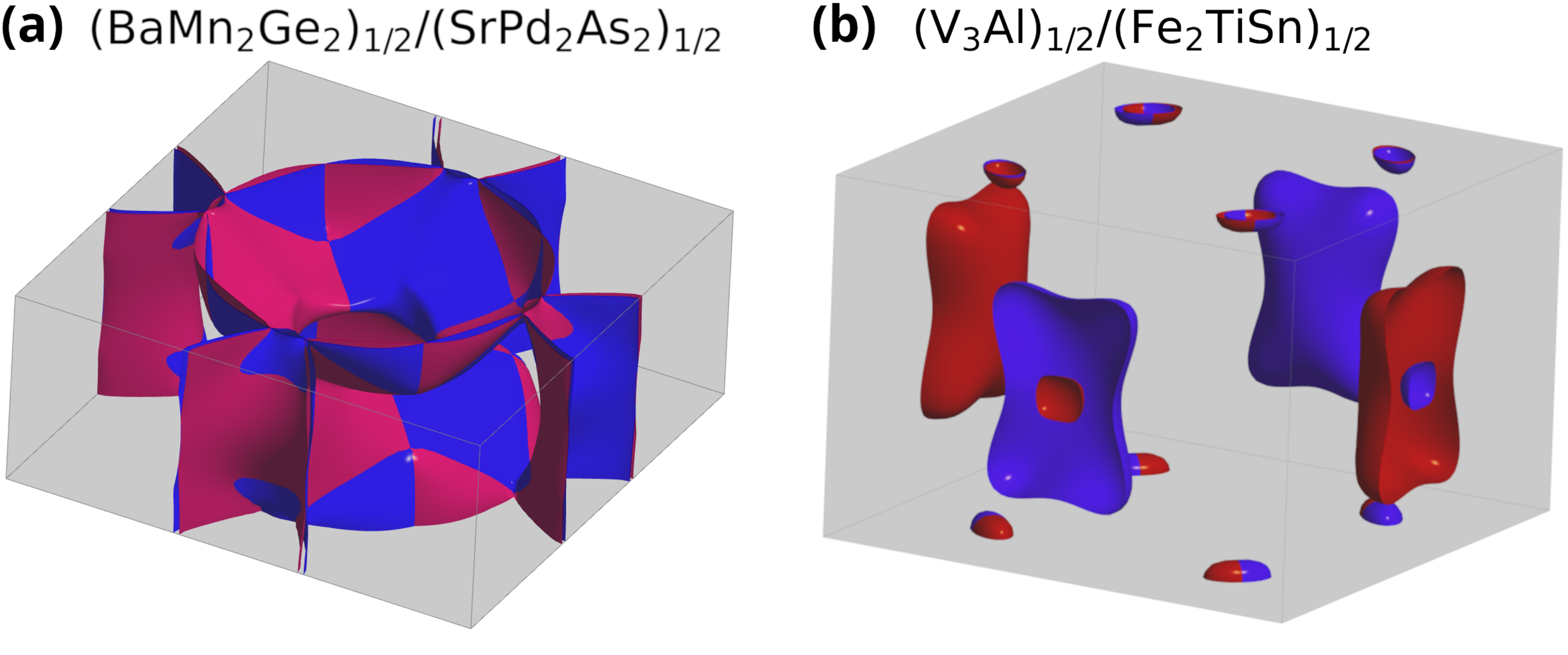}
\vspace{-0.4 cm}
\caption{\textbf{Spin-resolved Fermi surfaces of representative metallic $d$-wave altermagnetic superlattices.}
(a) The 122-based $(\mathrm{BaMn_2Ge_2})_{1/2}/(\mathrm{SrPd_2As_2})_{1/2}$ superlattice, for which a single relevant band crosses the Fermi level.
(b) The Heusler-based $(\mathrm{V_3Al})_{1/2}/(\mathrm{Fe_2TiSn})_{1/2}$ superlattice, showing the Fermi surfaces associated with the strongly spin-split band crossing near $X$ and $Y$. A second, weakly split hole-like pocket around $\Gamma$ is omitted for clarity. Blue and red surfaces denote opposite spin channels.}
\label{fig4}
\end{figure}

The electronic spin splitting is further illustrated by the spin-resolved Fermi surfaces of representative metallic superlattices. Figure~\ref{fig4}(a) shows the Fermi surfaces of the 122-based $(\mathrm{BaMn_2Ge_2})_{1/2}/(\mathrm{SrPd_2As_2})_{1/2}$ superlattice, where a single relevant band crosses the Fermi level and the opposite-spin sheets are separated over extended regions of momentum space. In the Heusler-based $(\mathrm{V_3Al})_{1/2}/(\mathrm{Fe_2TiSn})_{1/2}$ superlattice, two bands cross the Fermi level. A hole-like pocket around $\Gamma$ exhibits relatively weak splitting, whereas the band crossing near $X$ and $Y$ is more strongly split. Figure~\ref{fig4}(b) displays the Fermi surfaces of the latter band, omitting the weakly split $\Gamma$-centered pocket for clarity. In both systems, the momentum-dependent separation of opposite-spin Fermi-surface sheets differs from the approximately uniform exchange splitting characteristic of conventional ferromagnets.

Finally, we examine the influence of superlattice periodicity
and magnetic coupling on the spin splitting. Increasing the
constituent-layer thickness from
$(\mathrm{V_3Al})_{1/2}/(\mathrm{Fe_2TiSn})_{1/2}$
to $(\mathrm{V_3Al})_1/(\mathrm{Fe_2TiSn})_1$
reduces the maximum splitting from approximately
$206$ to $10$~meV [Figs.~\ref{fig3}(e,f)],
although both structures retain $P4mm$ (No.~99)
crystallographic symmetry. At the same time, the
magnetic coupling changes from C-type to G-type.
As shown in the Supplementary Information,
the reduction is not accompanied by a comparable
decrease in the local V magnetic moments.
The strong suppression of spin splitting may
therefore arise from changes in the electronic
structure associated with magnetic coupling
and constituent-layer thickness.

Our results identify semiconducting and metallic
superlattices with substantial nonrelativistic
$d$-wave spin splitting. The additional 122-based
and Heusler-based systems demonstrate the importance
of magnetic symmetry, while the variations with
constituent selection and superlattice periodicity
show that symmetry alone does not determine the
splitting magnitude. In the limit of thicker
constituent layers, the $d$-wave spin splitting
may become localized near the interfaces rather
than extending throughout the superlattice,
suggesting a possible connection between bulk
$d$-wave altermagnetism in short-period
superlattices and interface altermagnetism in
heterostructures of conventional antiferromagnets
or antiferromagnetic and nonmagnetic materials.
We next examine the magnetic stability and
collective spin excitations of the predicted
altermagnetic superlattices.

\subsection*{Magnetic stability and altermagnetic spin excitations}
\label{sec:magnetism}

Realizing altermagnetism above room temperature requires robust magnetic order. We therefore investigate the magnetic stability of the proposed superlattices through their exchange interactions, magnetocrystalline anisotropy, and N\'eel temperatures estimated within the random-phase approximation (RPA). All investigated altermagnetic superlattices exhibit an out-of-plane easy axis along [001], with magnetocrystalline anisotropy energies (MAEs) ranging from $0.11$ to $1.28$~meV (Table~\ref{tab:superlattices}). The calculated exchange interactions and computational details are provided in the Supplementary Information.

\begin{figure*}[!ht]
\centering
\includegraphics[width=1.0\textwidth]{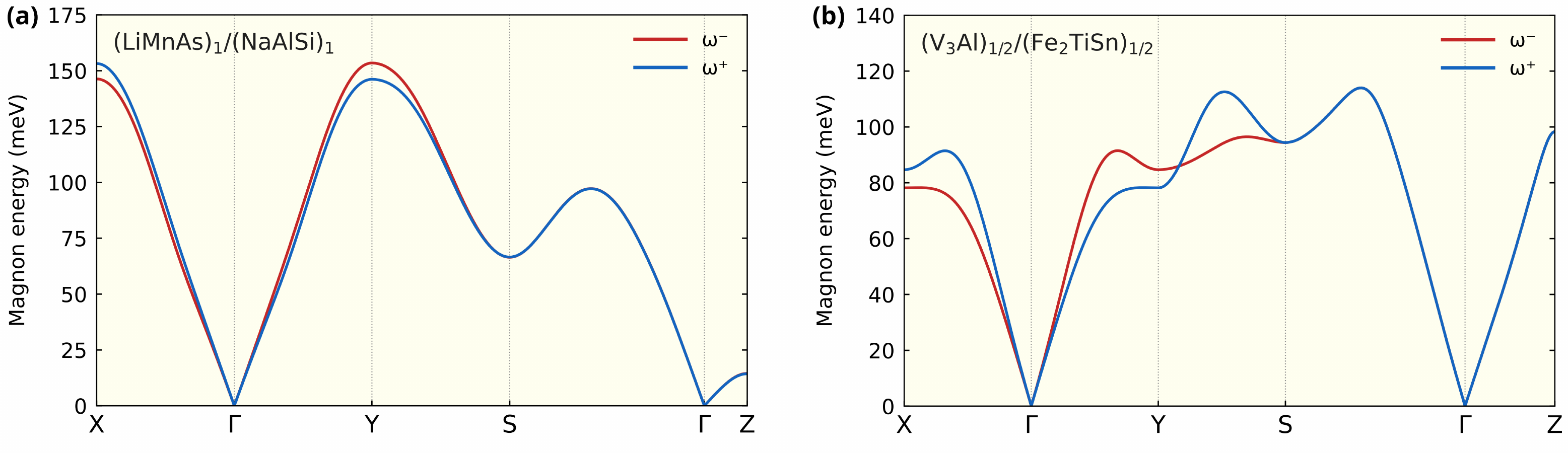}
\vspace{-0.8cm}
\caption{\textbf{Nonrelativistic $d$-wave magnon splitting in altermagnetic superlattices.}
Calculated magnon dispersions of (a) the 111-based $(\mathrm{LiMnAs})_1/(\mathrm{NaAlSi})_1$ and (b) the Heusler-based $(\mathrm{V_3Al})_{1/2}/(\mathrm{Fe_2TiSn})_{1/2}$ superlattices along the $X$--$\Gamma$--$Y$--$S$--$\Gamma$--$Z$ high-symmetry path. Red ($\omega^-$) and blue ($\omega^+$) curves represent oppositely polarized acoustic magnon branches obtained from first-principles exchange interactions within linear spin-wave theory, neglecting spin--orbit coupling and magnetic anisotropy. The reversal of their energy ordering between $X$--$\Gamma$ and $\Gamma$--$Y$ reflects the characteristic $d$-wave symmetry. The optical magnon branches of the Heusler-based superlattice are presented in the Supplementary Information.}
\label{fig5}
\end{figure*}

The calculated N\'eel temperatures (Table~\ref{tab:superlattices}) indicate that several superlattices can sustain magnetic order  above room temperature. For comparison, the parent compounds LiMnAs and NaMnAs have experimental N\'eel temperatures of approximately $374$ and $350$~K, respectively~\cite{Beleanu2013LiMnAs,Volny2022NaMnAs}. We employ a classical Heisenberg description to treat the different material families consistently, including systems with itinerant magnetic moments for which assigning fixed atomic spin quantum numbers is less straightforward. For the Mn-based 111 superlattices, the calculated Mn moments of approximately $4\,\mu_{\mathrm B}$ are consistent with an effective spin $S=2$. An approximate quantum-spin correction can be obtained by rescaling the classical RPA ordering temperature by $S(S+1)/S^2=1.5$~\cite{Sasioglu2005}. For the G-type $(\mathrm{LiMnAs})_1/(\mathrm{NaMnAs})_1$ superlattice, the classical RPA yields $T_{\mathrm N}\approx285$~K, corresponding to an approximate rescaled value of $428$~K. The classical RPA estimates for the 111-based $(\mathrm{LiMnAs})_1/(\mathrm{NaAlSi})_1$ and Heusler-based $(\mathrm{V_3Al})_{1/2}/(\mathrm{Fe_2TiSn})_{1/2}$ superlattices are $352$ and $325$~K, respectively.

The 122-based superlattices exhibit particularly strong magnetic ordering tendencies. For $(\mathrm{BaMn_2Ge_2})_{1/2}/(\mathrm{SrPd_2As_2})_{1/2}$, the classical RPA yields $T_{\mathrm N}\approx1423$~K. High ordering temperatures have also been observed experimentally in related Mn-based 122 compounds, including approximately $625$~K in BaMn$_2$As$_2$~\cite{Singh2009BaMn2As2} and $795$~K in BaMn$_2$P$_2$~\cite{Jacobs2023BaMn2P2}. Although the exchange interactions in the superlattice differ from those of the bulk compounds, these experimental results illustrate the strong antiferromagnetic interactions that can occur within the 122 structural family.

Beyond magnetic stability, collective spin excitations provide an additional signature of $d$-wave altermagnetism. Figure~\ref{fig5} presents the magnon dispersions of representative 111-based and Heusler-based superlattices, calculated from isotropic exchange interactions within linear spin-wave theory. Spin--orbit coupling and magnetic anisotropy are neglected to isolate the nonrelativistic contribution to the magnon splitting. The oppositely polarized acoustic magnon branches, $\omega^-$ and $\omega^+$, are degenerate at $\Gamma$ but split away from symmetry-enforced degeneracies. Their energy ordering reverses between $X$--$\Gamma$ and $\Gamma$--$Y$, reflecting the characteristic $d$-wave symmetry. The maximum splitting along the displayed path reaches approximately $7.3$~meV for $(\mathrm{LiMnAs})_1/(\mathrm{NaAlSi})_1$ and $18.7$~meV for $(\mathrm{V_3Al})_{1/2}/(\mathrm{Fe_2TiSn})_{1/2}$.

The $d$-wave magnon splitting arises from anisotropic intra-sublattice exchange interactions along orthogonal crystallographic directions, analogous to the anisotropic same-sublattice hopping responsible for electronic $d$-wave spin splitting in the effective tight-binding model. The exchange parameters and optical magnon branches of the Heusler-based superlattice are provided in the Supplementary Information. In metallic systems, coupling to itinerant Stoner excitations may produce magnon damping when the collective modes overlap with the particle--hole continuum, an effect not captured by the present Heisenberg description~\cite{Beida2025,Karlsson2000,Kotani2008,Lounis2010,Skovhus2021}. The calculated dispersions therefore describe exchange-driven magnon energies without accounting for finite magnon lifetimes. Together with the estimated ordering temperatures, these results indicate that robust antiferromagnetic order can coexist with characteristic nonrelativistic $d$-wave magnon splitting in the proposed superlattices.

\section*{Discussion}
\label{sec:discussion}

Our results demonstrate that periodic stacking of conventional spin-degenerate antiferromagnets with structurally compatible materials can give rise to bulk $d$-wave altermagnetism. The underlying mechanism is the removal of degeneracy-protecting symmetries while retaining the fourfold rotational symmetry of the constituent materials. Whether this symmetry reduction produces altermagnetism depends on the complete magnetic structure rather than the crystallographic space group alone. This distinction is illustrated by the Heusler-based superlattices, where constituent selection determines whether degeneracy-protecting symmetries are retained, and by the 122-based superlattices, where magnetic periodicity can restore spin degeneracy.

The generality of this design strategy is demonstrated by its applicability to three distinct structural families: 111-based, 122-based, and Heusler-based superlattices. Related proposals for SrCrO$_3$-based oxide heterostructures have explored mechanisms involving quantum confinement, orbital ordering, and structural distortions \cite{Bandyopadhyay2026}. The present results establish a broadly applicable symmetry-based approach to engineering bulk $d$-wave altermagnetism across structurally distinct material families.

Beyond its applicability to different material families, the superlattice approach offers opportunities to tune the magnitude of the electronic spin splitting through constituent selection and superlattice periodicity. Calculations for additional 122-based superlattices yield spin-splitting scales ranging from approximately $3$ to $177$~meV in chemical supercells, although global altermagnetic spin splitting is retained only for magnetic ground states with the required symmetry.
In the Heusler-based V$_3$Al/Fe$_2$TiSn superlattices, increasing the constituent
thickness, accompanied by a change in magnetic stacking from C-type to G-type,
reduces the splitting from 206 to 10 meV without changing the crystallographic
space group. These findings highlight the importance of material-specific electronic structure in determining the spin-splitting magnitude beyond the constraints imposed by symmetry.

The sensitivity to constituent selection also motivates exploring additional structural families, including 1111-type compounds. Superlattices combining materials from different families, such as 111/122, 111/1111, and 122/1111, could offer further opportunities to tune orbital hybridization, interfacial coordination, and magnetic exchange interactions. Such combinations would require suitable lattice matching, interfacial compatibility, and chemical stability. A systematic computational search could therefore combine structural and chemical screening with magnetic-symmetry analysis and first-principles evaluation of electronic and magnetic properties, including the stability of competing magnetic configurations.

Beyond achieving sizable electronic spin splitting, maintaining magnetic order above room temperature is essential for realizing high-temperature altermagnetism. Several superlattices investigated here exhibit calculated N\'eel temperatures in this range, together with out-of-plane magnetic easy axes and characteristic nonrelativistic $d$-wave magnon splitting. These findings indicate that altermagnetic electronic structures can coexist with robust antiferromagnetic order and distinctive collective spin excitations. The ordering-temperature estimates depend on the adopted Heisenberg and RPA descriptions, while the calculated magnon dispersions neglect damping from itinerant electronic excitations. Experimental verification of the magnetic ground states, ordering temperatures, and spin-resolved electronic structures will therefore be essential.

Overall, our findings establish a symmetry-based superlattice design strategy for realizing bulk $d$-wave altermagnetism across distinct structural families. The ability to influence electronic spin splitting through constituent selection and superlattice periodicity, combined with the predicted magnetic stability of several realizations, provides a foundation for the systematic discovery and optimization of room-temperature $d$-wave altermagnets.

\section*{Methods}

\subsection*{Tight-binding calculations}

To investigate the emergence of nonrelativistic
$d$-wave spin splitting, we employed the effective
tight-binding model introduced in
Eq.~(\ref{eq:single_layer_tb}).
The model describes two inequivalent layers,
each containing oppositely magnetized sublattices,
with direction-dependent intrasublattice hopping,
intersublattice hopping, and staggered exchange
fields. The electronic band structures were obtained
by diagonalizing the Hamiltonian in reciprocal space.
Model parameters and additional details are provided
in the Supplementary Information.

\subsection*{First-principles calculations}

First-principles calculations were performed
within density-functional theory using the
generalized gradient approximation in the
Perdew--Burke--Ernzerhof (PBE) parametrization,
as implemented in \textsc{QuantumATK}, with
norm-conserving PseudoDojo pseudopotentials
and localized atomic-orbital basis sets
~\cite{smidstrup2019an,perdew1996generalized,
QuantumATKPseudoDojo}.
The electronic band structures were calculated
within the collinear magnetic approximation
without spin--orbit coupling to identify
nonrelativistic altermagnetic spin splitting.
Atomic structures were relaxed until the
residual forces were below $0.01$~eV/\AA.
Three-dimensional periodic boundary conditions
were employed for all superlattices.
Magnetocrystalline anisotropy energies were
evaluated separately with spin--orbit coupling
included. Additional computational parameters,
including Brillouin-zone sampling and
structural optimization details, are provided
in the Supplementary Information.

\subsection*{Exchange interactions and magnetic excitations}

The interatomic Heisenberg exchange parameters $J_{ij}^{\alpha\beta}$ were calculated using the Liechtenstein--Katsnelson--Antropov--Gubanov (LKAG) method, based on the magnetic force theorem, as implemented in \textsc{QuantumATK}~\cite{Liechtenstein1987}. The resulting exchange interactions were mapped onto a multisublattice Heisenberg Hamiltonian and used to calculate the magnon dispersions within linear spin-wave theory using the \textsc{SpinW} package~\cite{Toth2015SpinW}. Spin--orbit coupling and magnetic anisotropy were neglected in the spin-wave calculations to isolate the nonrelativistic altermagnetic magnon splitting.

The N\'eel temperatures were estimated within the multisublattice random-phase approximation (RPA), following the Green's-function approach of Rusz \textit{et al.}~\cite{Rusz2005RPA,Rusz2006RPA}. A classical Heisenberg description was adopted consistently across the investigated superlattices. For the Mn-based systems with relatively large local magnetic moments, an approximate quantum-spin correction was additionally considered by rescaling the classical RPA ordering temperatures. Magnetocrystalline anisotropy was included in the ordering-temperature calculations.

\section*{Data availability}

The data supporting the findings of this study,
including the calculated electronic band structures,
exchange interactions, magnon dispersions, and
N\'eel temperatures, are available from the
corresponding author upon reasonable request.

\section*{Code availability}

First-principles calculations were performed using
the commercially available QuantumATK software,
while magnon dispersions were calculated using
the publicly available SpinW package.
The tight-binding Hamiltonian and model parameters
are provided in the Supplementary Information.
The custom codes used for the tight-binding
calculations and the random-phase approximation
(RPA) estimates of the N\'eel temperatures are
available from the corresponding author upon
reasonable request.

\section*{Acknowledgements}

This work was supported by the Deutsche
Forschungsgemeinschaft (DFG) through CRC/TRR 227,
Project No. B12 (Project ID 328545488), and
Grants No. LO 1659/10-1, No. LO 1659/13-1,
and No. SA 3174/1-1, and by the European
Innovation Council (EIC) Pathfinder Open project,
Grant No. 101129641.  The authors acknowledge the use of generative
AI tools to improve the clarity, readability,
and presentation of the manuscript.

\section*{Author contributions}

E.\c{S}. conceived the study and performed the
first-principles calculations.
B.G. performed the tight-binding calculations
and contributed to the theoretical analysis.
P.R.R. calculated the magnon dispersions and
estimated the N\'eel temperatures within the
random-phase approximation.
I.M. and S.L. supervised the project.
All authors contributed to the analysis and
interpretation of the results.
E.\c{S}. wrote the manuscript with input
from all authors.

\section*{Competing interests}

The authors declare no competing interests.

\bibliography{d-wave_superlattices}

\end{document}